\documentstyle[prd,aps]{revtex} 

\input psfig \pssilent
\begin{document}
\title{Canonical quantization of constrained theories on 
discrete space-time lattices}

\author{Cayetano Di Bartolo$^1$, Rodolfo Gambini$^2$, and Jorge Pullin$^3$}
\address{1. Departamento de F\'{\i}sica, Universidad Sim\'on Bol\'{\i}var,\\
Aptdo. 89000, Caracas 1080-A, Venezuela.}
\address{2. Instituto de F\'{\i}sica, Facultad de Ciencias, 
Universidad
de la Rep\'ublica\\ Igu\'a esq. Mataojo, CP 11400 Montevideo, Uruguay}
\address{3. Department of Physics and Astronomy, 
Louisiana State University,\\ 202 Nicholson Hall, Baton Rouge,
LA 70803-4001}

\date{May 29th 2002}
\maketitle
\begin{abstract}
We discuss the canonical quantization of systems formulated on
discrete space-times.  We start by analyzing the quantization of
simple mechanical systems with discrete time. The quantization becomes
challenging when the systems have anholonomic constraints. We propose
a new canonical formulation and quantization for such systems in terms
of discrete canonical transformations. This allows to construct, for
the first time, a canonical formulation for general constrained
mechanical systems with discrete time. We extend the analysis to gauge
field theories on the lattice. We consider a complete canonical
formulation, starting from a discrete action, for lattice Yang--Mills
theory discretized in space and Maxwell theory discretized in space
and time. After completing the treatment, the results can be shown to
coincide with the results of the traditional transfer matrix
method. We then apply the method to BF theory, yielding the first
lattice treatment for such a theory ever.  The framework presented
deals directly with the Lorentzian signature without requiring an
Euclidean rotation. The whole discussion is framed in such a way as to
provide a formalism that would allow a consistent, well defined,
canonical formulation and quantization of discrete general relativity,
which we will discuss in a forthcoming paper.

\end{abstract}

\section{Introduction}

Discretizations are used in general relativity mainly in three
contexts: a) numerical relativity, b) canonical quantum gravity, c)
path integral quantum gravity. In the first context the aim of
discretizing the equations is to solve them on a computer
numerically. In the second and third contexts the primary goal is to
make finite the infinite dimensional computations implied by the field
theory nature of general relativity (and eventually to solve the
theory numerically on a computer). It has long been known in numerical
relativity that the set of algebraic equations that results of
discretizing the Einstein equations (for instance, with externally
prescribed lapse and shift) is {\em inconsistent} in the
following sense. The discretized evolution equations produce solutions
in the future that do not satisfy the discretized constraints, even if
one started from initial data that satisfied them. The general
attitude towards this problem in numerical relativity has been that
the aim is to provide an approximate description of the continuum
theory and therefore the inconsistency is accepted as part of the 
errors of the approximation scheme. This was clearly discussed by Choptuik
\cite{choptuik}, who showed that the constraints were preserved at the
same order of discretization as the evolution equations.

In canonical and path integral quantum gravity, the potential
inconsistency of the discretized theories has not received a lot of
emphasis.  It was noted by Bander \cite{bander} in general and later
by Friedman and Jack \cite{FrJa} in the context of Regge Calculus that
it was not straightforward to find consistent Hamiltonian formulations
of general relativity discretized in space and with continuum time. It
has generally been recognized that a problem has to arise in lattice
treatments of general relativity since, unlike the case of Yang--Mills
theory, the introduction of a lattice breaks the ``gauge symmetry'' of
the theory (invariance under diffeomorphisms). In spite of these
observations, the remarkable formulation of a well defined Hamiltonian
constraint for quantum general relativity due to Thiemann \cite{Th}
starts with a discretization of the theory. Thiemann promotes the
Hamiltonian constraint to a quantum operator acting on the space of
solutions of the diffeomorphism constraint. This somewhat obscures the
issue of the implementation of the constraint algebra, since on such
space the algebra of constraints is  Abelian. It has been noted,
however, that it is likely that problems are implicit in the
quantization with respect to the closure of the constraint algebra
\cite{LeMa} ``off-shell'' (on non-diffeomorphism invariant
states). Proposals to construct quantizations of the Hamiltonian that
produce non-Abelian operators have failed to yield quantities that
reproduce the proper constraint algebra \cite{Sm,Ne}.

In canonical formulations, the role of the constraints is to generate
infinitesimal symmetries. In a discrete theory, the best we could hope
for is that the constraints generate some finite version of the
symmetry in question. Obviously, such an action cannot be achieved
with a discretization of the continuum constraints, since the latter
generate infinitesimal symmetries. The continuum constraints therefore
naturally structure themselves into algebras. The discrete generators
of symmetries that one could have in a discrete theory will not
structure themselves in an algebra. Therefore there is no sense in
which one will have in the discrete theory an algebra of
constraints that ``in the limit'' reproduces the continuum algebra.

An additional element is that discrete symmetries are not
characterized by free parameters, as infinitesimal symmetries are.  A
discrete transformation can only admit a set of fixed values for the
parameters of the transformation. Therefore in dealing with finite
symmetries the finite quantities that are remnants of the Lagrange
multipliers of the continuum theory become fixed. This inevitably
leads us to systems of constraints of second class in the canonical
treatment.

Finally, one has a choice as to the role of time. One could try to
discretize ``space'' and leave ``time'' continuous. This is highly
unnatural in general relativity and might further complicate the
issues of the constraint algebra since the latter mixes symmetries in
space and time. In this paper we will choose to pick a discrete time
in the theories we consider. This is more symmetric and allows for a
better connection with ``spin-foam'' approaches in which time is
discrete. 

Having a discrete canonical formulation for general relativity is,
in our opinion, fundamental at the time of understanding current
``spin-foam'' approaches to the path integral of quantum general
relativity. It is well known that path integrals for gauge field
theories should be evaluated with care in order to avoid 
integrating over the orbits of the gauge groups. This is at the
core of the Fadeev--Popov technique and can only be fully 
understood if one understands the degrees of freedom of the theory
at a Hamiltonian level \cite{teitel}. 
To achieve this for discrete gravity is
one of the goals we are pursuing.

An additional point to consider is that most lattice treatments 
of gauge theories require an Euclidean rotation in order to be
well defined. This is expected to be problematic in the case of
general relativity. The formalism we will present in this paper
does not require an Euclidean rotation and operates directly 
in terms of the Lorentzian signature theory. The lack of a Lorentzian
formulation on the lattice has been recognized as one of the major
unsolved issues in lattice formulations of gravity \cite{Loll}

The inconsistencies that arise in discretizations are present in spin
foam approaches directly. When one discretizes the action in order to
compute the path integral one is left with a discrete action whose
equations of motion are seldom analyzed and in general yield an
inconsistent theory. If one therefore computes the integral for the
discrete theory, one is attempting a functional integration for a 
theory that does not exist.  One will get a result, but there will
generically be no way of making a connection with the underlying
classical theory.

A point to consider is if one could not simply adopt in the quantum
context the same attitude as was adopted in numerical relativity.
Could one deal with inconsistent discrete theories that are just
approximations of a consistent theory that exists in the limit?  This
would be a point of view that is tenable if one had a well defined way
of taking the limit in which discretizations are infinitely
refined. This is not currently available for general relativity and it
is generally thought that such limit will be highly non-trivial given
the non-renormalizability of the theory.  Current thinking favors
viewing the discrete theory as existing at a fundamental level and
never taking the limit (see for instance \cite{kauffman}. In such a
case one needs to take the issue of consistency of such a theory
seriously. Even if one wished to consider some sort of ``continuum
limit'' (for instance summing over all possible discretizations), it
is our belief that one has a better chance of defining such a limit if
one is obtaining it in a process where one considers a sequence of
well defined, consistent, theories.

In this paper we would like to discuss the canonical treatment of
discrete systems. The subject is quite non-trivial. Up to this paper
the state of the art of the subject was such that even the treatment
of simple mechanical systems with constraints was incomplete.  In
particular, the literature only covers systems with constraints that
are holonomic (function of half of the variables of the phase
space). In this paper we extend the canonical formulation to include
arbitrary constrained mechanical systems. A key new ingredient is to
introduce a symplectic structure on the lattice based on discrete
canonical transformations that implement the discrete time evolution.
We then turn our attention to the treatment of gauge field theories on
the lattice. To our knowledge, this is the first time that a complete
canonical analysis of gauge field theories discretized on the lattice
is presented, in the sense that we discretize the action of the theory
and derive the canonical theory from the discrete action. The final
results for Maxwell and Yang--Mills theories on the lattice
coincides with the traditional transfer matrix methods, but the 
intermediate steps are quite non-trivial. We also present a canonical
treatment of BF theory on the lattice. In this case, up to present
no lattice treatment was present at all in the literature. The 
formulation we construct is readily applicable to the case of 
general relativity, which we will discuss in a separate publication.

This paper is organized as follows. In section II we discuss the
quantization of mechanical systems with discrete time evolution.  In
section III we extend the analysis to lattice Yang--Mills theory.  In
section IV we discuss discretized BF theory. We end with a discussion
of implications for quantum gravity.

\section{Mechanics with discrete time}
\subsection{Summary of previous results}
Some of the difficulties of constructing Hamiltonian formulations for
discrete theories can be illustrated with the simplest mechanical
systems if we assume time is a discrete variable. The subject of
``discrete mechanics'' has been considered several times in the
literature. The broad subject of ``symplectic integrators'' also
touches upon these issues. However, all discussions we have found in
the literature refer to systems with holonomic constraints, for
reasons that will soon become apparent \cite{Baez}. Here we will
present a a discussion that includes anholonomic constraints, as the
ones in gravity.

Discretizing time in mechanics immediately leads to difficulties
with conservation laws. Let us start by considering the action of 
a particle of unit mass 
in a potential, written in a discrete approximation 
in which time is split into intervals of duration $\epsilon$,
\begin{equation}
S=\sum_{i=1}^{N} p_i (q_{i+1}-q_i) + 
\left[{p_i}^2/2 +V(q_i)\right]\epsilon.
\end{equation}
If we work out the equations of motion of this action, one gets,
\begin{equation}
q_{i+1} -q_i = \epsilon p_i, \qquad p_{i}-p_{i-1}=-V'(q_i) \epsilon.
\end{equation}
One immediately recognizes a discrete approximation to Hamilton's
equation. There is an asymmetry in the equations in the sense that the
discretization of the time derivative of $q$ is ``centered'' between
$i$ and $i+1$ whereas the one in $p$ is centered between $i-1$ and
$i$. This theory recovers in the continuum limit the usual classical mechanics
of a particle in a potential. However, it is well known that in the
discrete theory, energy is not conserved. 
It is immediate to realize that similar
problems will arise in constrained systems: the discrete evolution
equations will fail to preserve the constraints, at least if they are
anholonomic.

A first attempt to solve the problem of non-conservation is due
to T.D. Lee \cite{TDLee}. His proposal consists in ``parameterizing''
the system by including the time interval as a new variable. The
action can therefore be written as,
\begin{equation}
S = \sum_n p_n (q_{n+1}-q_n) -H(p_n,q_n)(t_{n+1}-t_n)
\end{equation}
and we consider as variables $p_i,q_i$ and $t_i$. When
one varies with respect to $t_i$ one immediately gets that
$H(p_i,q_i)=H(p_{i-1},q_{i-1})$. This equation, with the 
two other equations of motion determine $q_i,p_i$ and $t_i$, 
that is, the time interval is now determined by the equations
of motion. 

T.D. Lee's construction merits several comments. First, notice
that the use of the parameterization immediately suggests how
to handle the problem of non preservation of constraints in
general relativity. The idea is to consider the Einstein 
evolution equations together with the constraints as equations
for the metric, extrinsic curvature {\em and} the Lagrange
multipliers (the lapse and shift). One has therefore four 
additional equations (the constraints) and four additional 
variables. One can therefore construct a discrete system of
equations that preserves the constraints in time. We are
currently studying the feasibility of implementing this scheme
in a numerical simulation of spherically symmetric space-times.

There has been considerable discussion in the literature on
constructing quantities that are kept invariant by the 
evolution implied by the discretized evolution equations.
Maeda \cite{Maeda} constructed a modified Noether theorem.
Logan \cite{Logan} proposed a method for constructing invariants
not associated with symmetries of the action.
Jaroszkiewicz and Norton \cite{JaNo} discussed the quantization
of systems by constructing evolution operators. 

A whole separate chapter could be devoted to the subject of
``symplectic integrators'' \cite{symp}. The focus of this area of
research is usually numerical. People have noticed for a long time
that it is desirable for discretization of dynamical equations to
preserve certain conserved quantities of the classical theory. In
particular the symplectic structure, or, as in the case of celestial
dynamics, angular momentum and energy \cite{pretre}.  This is a highly
developed area of research, in which canonical techniques have been
widely used in the treatment of discrete systems. In fact, our current
paper can be viewed as an extension of certain symplectic integration
results to the realm of constrained systems, with an eye towards
application to general relativity.

\subsection{Canonical formulation and quantization 
of simple discrete mechanical  systems}

The classical canonical treatment of discrete systems has 
been considered in the literature (see for instance \cite{pretre}). 
The usual treatment considers unconstrained systems and implements
time evolution through 
canonical transformations of type 2 \cite{goldstein}. The discrete 
evolution produces time histories of the variables. In the continuum
limit, some of these evolutions (the ones with good continuum limit) 
will be solutions of the continuum equations of motion. 
Since we are interested in constrained systems, we will find that 
it is more convenient to work with canonical transformations of type 1.

%
%

We start by denoting $\hat{L}(q,\dot{q})$ the Lagrangian of the
continuum theory. We discretize time in equal intervals
$t_{n+1}-t_n=\varepsilon$ and we label the generalized coordinates
evaluated at time $t_n$ as $q_n$.
We define the ``discretized'' Lagrangian as,
\begin{equation}
L(n,n+1) \equiv L(q_n,q_{n+1})\equiv\varepsilon \hat{L}(q,\dot{q})
\end{equation}
where
\begin{equation}
q=q_n \qquad\mbox{and}\qquad  \dot{q} \equiv
\frac{q_{n+1}-q_n}{\varepsilon}.
\end{equation}
The action can then be written as,
\begin{equation}
S= \sum_{n=0}^N L(q_n,q_{n+1})
\end{equation}
and its Lagrange equations of motion,
\begin{equation}
{\partial S \over \partial q_n} = {\partial L(q_{n-1},q_{n}) \over 
\partial
q_n}+{\partial L(q_{n},q_{n+1}) \over \partial q_n} =0. 
\end{equation}
As can be seen, in the discrete theory the Lagrangian is not a
function of the manifold of $q$'s and its tangent bundle ${\dot q}$,
but it is a function of $q_n$ at the $n-th$ level and of $q_{n+1}$ at
the next level (or equivalently $q_n$ and $q_{n-1}$).

We now wish to introduce canonically conjugate momenta and show how
the discrete evolution can be represented as a canonical transformation
in phase space. The partial derivatives of $L$ and $\hat{L}$ 
are related by,
\begin{eqnarray}
\frac{\partial L(n,n+1)}{\partial q_{n+1}} &=& \frac{\partial
\hat{L}}{\partial \dot{q}} \label{der1}\\
\frac{\partial L(n,n+1)}{\partial q_n} &=&
\varepsilon\frac{\partial \hat{L}}{\partial q} -\frac{\partial
\hat{L}}{\partial \dot{q}}\label{der2}
\end{eqnarray}

In the continuum, the momentum canonically conjugate to $q$ is 
defined as
\begin{equation}\label{pcontinuo}
p=\frac{\partial\hat{L}}{\partial\dot{q}}
\end{equation}
and due to (\ref{der1})  we define the canonically conjugate momentum
in the discrete theory as
\begin{equation}\label{P1}
p_{n+1} = \frac{\partial L(n,n+1)}{\partial q_{n+1}}.
\end{equation}

The Lagrange equations in the continuum can be written as,
\begin{equation}\label{Clagrange}
\frac{d\,p}{dt} = \frac{\partial \hat{L}}{\partial q}
\end{equation}
with $p$ defined as in (\ref{pcontinuo}). 
To discretize this last expression we make the substitution
\begin{equation}
\dot{p} \rightarrow \frac{p_{n+1}-p_n}{\varepsilon}
\end{equation}
and use (\ref{der1}), (\ref{der2}) and (\ref{P1}) to get
\[
\frac{p_{n+1}-p_n}{\varepsilon} = \frac{\partial \hat{L}}{\partial
q} =\frac{1}{\varepsilon}\left[\frac{\partial \hat{L}}{\partial
\dot{q}} +\frac{\partial L(n,n+1)}{\partial q_n}\right]
=\frac{1}{\varepsilon}\left[ \frac{\partial L(n,n+1)}{\partial
q_{n+1}}  +\frac{\partial L(n,n+1)}{\partial q_n}\right] =
\frac{1}{\varepsilon}\left[ p_{n+1}  +\frac{\partial
L(n,n+1)}{\partial q_n}\right],
\]
Then
\begin{equation}\label{P2}
  p_n = - \frac{\partial
L(n,n+1)}{\partial q_n}.
\end{equation}
Summarizing, the discrete Lagrange equations are,
\begin{equation}
p_{n+1} = \frac{\partial L(q_n,q_{n+1})}{\partial q_{n+1}}
\qquad,\qquad p_n = - \frac{\partial L(q_n,q_{n+1})}{\partial q_n}
\,.\label{15}
\end{equation}
And it should be noted that these equations define a type 1 canonical
transformation from the variables
$(q_n,p_n)$ to $(q_{n+1},p_{n+1})$.

The variables $(q_n,p_n)$ 
constitute a phase space and as usual, if the system of equations 
(\ref{15}) is non-singular one can introduce new coordinates
$(q_{n+1},p_{n+1})$ for the phase space. This transformation is 
canonical, in the sense that it preserves the symplectic structure,
\begin{equation}
\left\{q_n,p_n\right\}=\left\{q_{n+1},p_{n+1}\right\}=1.
\end{equation}

The generating function (we use the terminology of \cite {saletan}) 
for the transformation is $F(q_n,p_n)=-L(q_n,q_{n+1}(q_n,p_n))$,
where $q_{n+1}(q_n,p_n)$ is given by inverting the first  of (\ref{15});
it is immediate to check that it satisfies the equations of a generating
function,
\begin{eqnarray}
{\partial F \over \partial q_n} &=&p_n -{\partial q_{n+1}\over \partial q_n}p_{n+1}\\
{\partial F \over \partial p_n} &=& -{\partial q_{n+1}\over \partial p_n}p_{n+1}.
\end{eqnarray}

One can now introduce the generating functions of canonical transformations
of the various types that
correspond to discrete time evolution. If the variables
$(q_n,q_{n+1})$ are used for coordinates of the phase space, then one
can introduce a generating function of type 1,
\begin{eqnarray}
F_1(q_n,q_{n+1})&\equiv& -L(q_n,q_{n+1})\\
p_n &=& {\partial F_1\over \partial q_n} =\left\{F_1,p_n\right\}\\
p_{n+1}&=&-{\partial F_1\over \partial q_{n+1}} =-\left\{F_1,p_{n+1}\right\}.
\end{eqnarray}

It is then possible to define generating functions of type 2,3 and
4. For instance, if the phase space is coordinatized by
$(q_n,p_{n+1})$, solving the first of (\ref{15}) 
for $q_{n+1}$ one can introduce a
generating function of type 2,
\begin{eqnarray}
F_2(q_n,p_{n+1})&\equiv& q_{n+1} p_{n+1} -L(q_n,q_{n+1}),\\
p_n&=&{\partial F_2\over \partial q_n} =\left\{F_2,p_n\right\},\label{12}\\
q_{n+1}&=& {\partial F_2 \over \partial p_{n+1}} =
-\left\{F_2,q_{n+1}\right\}.\label{13}
\end{eqnarray}

To facilitate the comparison with the continuum limit case, we
can introduce a function $H(p_{n+1},q_{n})$ given by,
\begin{equation}
F_2(q_{n},p_{n+1})=p_{n+1} q_{n} +H(p_{n+1},q_{n}),
\end{equation}
such that one recovers the discrete Hamilton equations,
\begin{eqnarray}
q_{n+1} &=& q_{n} +{\partial H \over \partial p_{n+1}}\\
p_{n} &=& p_{n+1} +{\partial H \over \partial q_{n}}.
\end{eqnarray}

The quantization of these finite canonical systems is achieved by
representing the finite canonical transformations that materialize
the time evolution as unitary operators. We illustrate this with
a simple example, a particle in a potential. The Lagrangian is
given by
\begin{equation}
L(q_{n},q_{n+1})= m {(q_{n+1}-q_{n})^2 \over 2 \epsilon} - V(q_{n})
\epsilon 
\end{equation}
the canonical momentum is given by $p_{n+1} = m
(q_{n+1} -q_{n})/\epsilon$, from which we can get $q_{n+1}=p_{n+1} \epsilon/m
+q_{n}$. The generating function is given by,
\begin{equation}
F_2=p_{n+1} q_{n} + {p_{n+1}^2 \over 2 m} \epsilon +V(q_{n+1}) \epsilon =
p_{n+1} q_{n} +H(p_{n+1}, q_{n}),
\end{equation}
and the equations  of motion (taking into account (\ref{12},\ref{13})) are
\begin{eqnarray}
q_{n+1}&=&q_{n}+{p_{n+1} \over m} \epsilon\\
p_{n} &=& p_{n+1} + V'(q_{n}) \epsilon
\end{eqnarray}
which can be solved for $p_{n+1}$ as,
\begin{eqnarray}
q_{n+1} &=& q_{n}+ {p_{n} \over m} \epsilon -V'(q_{n}) 
 {\epsilon^2\over m}\\
p_{n+1} &=& p_{n} - V'(q_{n}) 
\epsilon. 
\end{eqnarray}

We now proceed to quantize the system. We choose a polarization such
that the wavefunctions are functions of the configuration variables,
$\Psi(q_{n+1})$. The canonical operators have the usual form. The
evolution of the system is implemented via a unitary
transformation. An immediate computation shows that the operator that
implements the transformation $p_{n+1} = U p_{n} U^\dagger$, 
$q_{n+1}=U q_{n} U^\dagger$ is given by,
\begin{equation}
U= \exp\left(i{V(q_{n})\epsilon\over \hbar}\right)
\exp\left(i{p_{n}^2\epsilon\over 2m \hbar}\right).
\end{equation}

At a quantum mechanical level the energy $H^0(q_{n+1},p_{n+1})$ is not
conserved, as we expected from the fact that it was not conserved
classically. It is remarkable however, that one can construct an
``energy'' (both at a quantum mechanical and classical level) that is
conserved by the discrete evolution. To construct this energy we use
the Baker--Campbell--Hausdorff formula,
\begin{equation}
\exp\left(X\right) \exp\left(Y\right)= \exp\left(X+Y+{1 \over 2}
\left[X,Y\right]+ {1 \over 12}\left(\left[X,\left[X,Y\right]\right]+
\left[Y,\left[Y,X\right]\right]\right)+\ldots\right),
\end{equation}
and one can therefore write $U=\exp\left({i\epsilon \over \hbar}
H_{\rm eff}(q_i,p_i)\right)$ where $H_{\rm
eff}=H^0(q,p)+O\left(\epsilon^2\right)$, which is immediately
conserved under evolution. It is straightforward to write down
a classical counterpart of this expression. 

\subsection{Canonical formulation and quantization 
of constrained discrete mechanical  systems}

We now extend the analysis of the previous subsection to the case of a
constrained system. Here it will become clear the advantage of using
canonical transformations of type 1. We start with an action written in
first order form,
\begin{equation}
L(n,n+1) = p_n (q_{n+1}-q_n) -\epsilon H(q_n,p_n)-\lambda_{nB} \phi^B(q_n,p_n)
\end{equation}
where we assume we have $M$ constraints $B=1\ldots M$.

We now exhibit the construction of a type 1 canonical transformation. We
construct the appropriate canonically conjugate momenta using the first
of (\ref{15}),
\begin{eqnarray}
P^q_{n+1} &=& {\partial L(n,n+1) \over \partial q_{n+1}} =p_n\label{37}\\
P^p_{n+1}&=& {\partial L(n,n+1) \over \partial p_{n+1}}= 0\label{38}\\
P^{\lambda_B}_{n+1}&=& {\partial L(n,n+1) \over \partial \lambda_{(n+1)B}}= 0.\label{49}
\end{eqnarray}

To determine the equations of motion for the system we start from the second set of equations (\ref{15}),
\begin{eqnarray}
P^q_n &=& -{\partial L(n,n+1) 
\over \partial q_{n}} =p_n+\epsilon {\partial H(q_n,p_n) \over \partial q_n} + \lambda_{nB}
{\partial \phi^B(q_n,p_n) \over \partial q_n}\label{40}\\
P^p_n &=& -{\partial L(n,n+1) 
\over \partial p_{n}} = -(q_{n+1}-q_n) + \epsilon {\partial H(q_n,p_n) \over \partial p_n} 
+ \lambda_{nB}
{\partial \phi^B(q_n,p_n) \over \partial p_n}\label{41}\\
P^{\lambda_B}_n &=& \phi^B(q_n,p_n).\label{42}
\end{eqnarray}

Combining the last two sets of equations we get the equations of motion for the system,
\begin{eqnarray}
p_n-p_{n-1} &=& -\epsilon {\partial H(q_n,p_n) \over \partial q_n} - \lambda_{nB}
{\partial \phi^B(q_n,p_n) \over \partial q_n}\label{43}\\
q_{n+1}-q_n &=&
\epsilon {\partial H(q_n,p_n) \over \partial p_n} 
+ \lambda_{nB} {\partial \phi^B(q_n,p_n) \over \partial p_n}\label{44}\\
\phi^B(q_n,p_n) &=&0.\label{45}
\end{eqnarray}

Superficially, these equations appear entirely equivalent to the continuum ones.
However, they hide the fact that in order for the constraints to be preserved,
the Lagrange multipliers get fixed. Another way to see it, is that $
P^q_{n+1} =p_n$ and therefore it is immediate that the Poisson bracket of the
constraints evaluated at $n$ and at $n+1$ is non-vanishing. We then consider 
the constraint equation (\ref{45}) and substitute $p_n$ by (\ref{37}),
\begin{equation}
\phi^B(q_n,P^q_{n+1})=0.
\end{equation}
We then solve (\ref{44}) for $q_n$ and substitute it in the previous equation,
one gets, 
\begin{equation}
\phi^B(q_{n+1},P^q_{n+1},\lambda_{nB})=0, \label{system}
\end{equation}
and this constitutes a system of equations. Generically, these will 
determine 
\begin{equation}
\lambda_{nB} =\lambda_{nB}(q_{n+1},P^q_{n+1}, v^\alpha) \label{lambdadet}
\end{equation}
where
the $v^\alpha$ are a set of free parameters that may arise if the system of
equations is undetermined. The eventual presence of these parameters will signify that
the resulting theory still has a genuine gauge freedom represented by freely
specifiable Lagrange multipliers. 

The final set of evolution equations for the system is therefore given
by (\ref{43},\ref{44}) where the Lagrange multipliers are substituted
using (\ref{lambdadet}).

At this point it is worthwhile discussing the physical meaning of the
determination of the value of the Lagrange multipliers. This is an issue
that is potentially confusing. Lagrange multipliers are normally associated
with gauge symmetries. The fixation of the value of the 
Lagrange multipliers seems to suggest that the constructive procedure of
the discrete theory somehow has selected a preferred gauge for us. A priori
this might appear very unnatural. How could the procedure know which gauge
to choose? A point to emphasize is that in this section we have 
absorbed in the definition of the Lagrange multipliers a factor of $\epsilon$
(the evolution parameter interval) when we replaced the integral of 
the action by a discrete sum. What gets determined is therefore
``$\lambda\times \epsilon$''. For a completely parameterized theory, since
there is no explicit reference to $\epsilon$, one can choose it 
arbitrarily and therefore redefine the value of the multiplier arbitrarily.
A different way of putting this, which might help later understand the
situation in general relativity, is to consider that once the lattice has
been established, the gauge is fixed, but one has still the freedom of
where to place the lattice in an arbitrary manner.

\section{Yang--Mills and Maxwell theories on a discrete space-time}

Having set up the canonical quantization of mechanical systems
discretized in time, we now turn our attention to field theories.
In this case we also wish to discretize space. Although the words
``Hamiltonian formulation'' are used frequently in the lattice 
context, the formulation is not constructed in the traditional
canonical sense. A popular way of obtaining a Hamiltonian operator
for quantized lattice gauge theory is the ``transfer matrix'' method.
This consists in discretizing the space-time action and 
reading off from the path integral the operator that would correspond
to the infinitesimal evolution operator in the limit in which the
spacing in time goes to zero. There is no construction of a classical
discrete theory that is later quantized. Even the definition of
canonical variables for such a classical theory is problematic
and has received some attention \cite{ReSm}. What we will do in
this paper is precisely that: we will construct a classical 
canonical discrete lattice theory for Yang--Mills, that we will
quantize and we will show that we recover the same results as
those of the usual transfer matrix approach.

We will first explore the construction of Yang--Mills lattice theory
with a continuum time parameter. Although we know that ultimately this
is not what we want, since as we argued it is unlikely a formulation
like this will be useful in the gravitational case, it will prove
instructive for later constructing a theory discretized in space and
time. We will then consider the quantization of Maxwell theory on 
a discrete space-time. Maxwell's theory has all the ingredients of
interest and is simpler than Yang--Mills theory.

\subsection{Discrete lattice Yang--Mills with continuum time}

We consider a hypercubic lattice in four dimensions with spatial
spacing $a$ and spacing in time $a_0$. We label the vertices of the
hypercube with an integer $i$.  Each link connecting two sites is
identified as a pair $\{i,j\}$. Associated with each (oriented) link
there is a holonomy $U_{ij}$, we shall assume for simplicity that it
takes value in $SU(2)$. The holonomy of the link traversed in the
opposite direction is given by the inverse $U_{ji}=U_{ij}^{-1}$. The
Wilson action is given by,
\begin{equation}
S= -{a_0 a^3\over 2} \sum_{P} {S_P \over A_P^2}{\rm Tr}\left(U_{P}\right)
\end{equation}
where the sum is over all elementary plaquettes $P$ and $A_P$ is the 
area of the plaquettes. $S_P$ is $-1$ if the plaquette is spatial 
and $+1$ if it is timelike. Notice that we are considering the Lorentzian
action, our treatment does not require a Wick rotation to the
Euclidean action. We now will take
the limit in which the spacing in the time direction, $a_0$ goes to
zero. In order to do this we will relabel the links with two
indices $l,n$ where the index $n$ labels the position of the vertex
``in time'' and the index $l$ runs through all the oriented links that
form the spatial sub-lattice $L_n$ at a
given time $n$. We denote by $\bullet l$ and $l\bullet$ the vertices
at the origin and end of the link $l$. We rewrite the action and 
we make explicit the 
elementary plaquettes that are completely spatial and those that
have temporal links,
\begin{eqnarray}
S &=& -{a_0 a^3\over 2} \sum_{n}\left[ {1 \over (a_0 a)^2} \sum_{l \in L_n}  {\rm Tr}\left(
U^{-1}_{l,n+1} V_{\bullet l,n}^{-1} U_{l,n} V_{l\bullet,n}\right)
-{1 \over a^4} \sum_{P_n} {\rm Tr}(U_{P_n})\right] 
\nonumber
\end{eqnarray}
where holonomies along timelike links are denoted by the spatial
label of the vertex they start and end at, and the
time level they start at, like $V_{l\bullet,n}$. 
The second term in the Lagrangian is a sum over all spatial
elementary plaquettes at time $n$, $P_n$.  We introduced the
following notation for the components of the holonomy, 
$U_{l} = U^I_{l} T^I$ where $U^I_{l}$ are real, the $T^I$ are given by 
$T^0=I/\sqrt{2}$, $T^a=-i \sigma^a/\sqrt{2}, a=1\dots3$ 
and $\sigma$ are 
the Pauli matrices. 

We will now take the limit in which $a_0\rightarrow 0$. The 
holonomies along the timelike direction become,
\begin{eqnarray}
V_{v,n} &=& 1 + \lambda_v a_0 + {(\lambda_v)^2\over 2} a_0^2 +\ldots\\
V^{-1}_{v,n} &=&
1 - \lambda_v a_0 + (\lambda_v)^2 {a_0^2\over 2} +\ldots\\
U_{l,n+1} &=& U_{l,n} + a_0 \dot{U}_{l,n} + {a_0^2\over 2}
\ddot{U}_{l,n}\ldots
\end{eqnarray}
where $v$ is any vertex of $L_n$ and $\lambda_v =\lambda^a_v T^a$ 
with $\lambda^a_v$ real.
We will use as generalized coordinates the variables $U^I_l$ and $\lambda^I_v$, normalizing
the Lagrangian in such a way that $S=\sum_n a_0 L(n,n+1)$

The Lagrangian therefore becomes,
\begin{eqnarray}
L &=& a  \sum_{|l|\in L_n} {\rm Tr}\left[ \dot{U}^{-1}_{l}
\lambda_{\bullet l} U_{l}+U^{-1}_{l} \dot{U}_{l}
\lambda_{l\bullet} + \lambda_{l\bullet} U^{-1}_{l}
\lambda_{\bullet l} U_{l} + {1 \over 2} \dot{U}^{-1}_{l}
\dot{U}_{l} -{\lambda_{\bullet l}^2 \over 2}
-{\lambda_{l\bullet}^2 \over 2} \right]
 \nonumber\\
&&+{1 \over 2a}  \sum_{P} {\rm Tr}\left(U_{P}\right) + {a^3 }
\sum_{l} \alpha_{l}\sum_{I=0}^3 \left[ \left(U_{l}^I\right)^2
-2\right]\label{70}
\end{eqnarray}

We have dropped the subscript $n$ since all variables are now 
evaluated on the same time slice. The above sum goes over
all links traversed only once in a given direction and 
$U_{\overline{l}}=U^{-1}_l$. We have also included an additional term
in the Lagrangian, which is a constraint that fixes the holonomies to
take values in $SU(2)$ provided the $\lambda^I_v$'s are real.

It can be seen that the discrete Lagrange equations of motion coming
from this Lagrangian lead, in the limit $a\rightarrow 0$ to
Yang--Mills equations. The discretization in this case does not
break any gauge symmetry and therefore the discretized theory is
automatically consistent.

We now proceed to work out the canonically conjugate momenta,
\begin{eqnarray}
\Pi^U_{l}\equiv  \Pi^{UI}_l T^I &=& {\partial L \over \partial
\dot{U}^0_l}T^0- {\partial L \over \partial \dot{U}^a_l}T^a=
a\left( \dot{U}^{-1}_{l} + \lambda_{l\bullet} U^{-1}_l - U^{-1}_l
\lambda_{\bullet l}
\right)\\
\Pi^\alpha_l &=& {\partial L \over \partial
\dot{\alpha}_l}  = 0 \label{vin1}\\
\left(\Pi^\lambda_v\right)^a &=& {\partial L \over \partial
\dot{\lambda}_v^a} = 0\label{vin2}
\end{eqnarray}

Performing the Legendre transform, the Hamiltonian becomes,
\begin{eqnarray}
H &=& \sum_{|l|\in L_n} {\rm Tr} \left[\frac{1}{2a} \Pi^U_{l}
\Pi^U_{\overline{l}} + \Pi^U_{l}(U_l\lambda_{l\bullet} -
\lambda_{\bullet l} U_l) \right] -{1 \over 2a}  \sum_{P_n} {\rm
Tr}\left(U_{P_n}\right) \nonumber
\\
&& - {a^3 } \sum_{l} \alpha_{l} \sum_{I=0}^3\left[
\left(U_{l}^I\right)^2 -2\right] +\sum_{|l|\in L_n} \rho_{l}
\Pi^\alpha_{l} + \sum_v \sum_{a=1}^3 \mu_{va}
\left(\Pi^\lambda_v\right)^a\label{74}
\end{eqnarray}
where $\rho_l$ and $\mu_{va}$ are Lagrange multipliers. We now need to
ensure that the evolution preserves the constraints. Ensuring the
conservation of (\ref{vin1}) leads to 
\begin{equation}
\sum_{I=0}^3 \left[ \left(U_{l}^I\right)^2 -2\right]=0
\label{vin3}
\end{equation}
and conservation of (\ref{vin2}) leads to Gauss' law, as we will
soon discuss. We now need to ensure the conservation of (\ref{vin3})
and of Gauss' law. Conservation of (\ref{vin3}) leads to 
\begin{equation}
{\rm Tr}\left(U_l \Pi^U_l\right)=0.\label{vin4}
\end{equation}
This implies that the product $E_{l}\equiv U_{l}
\Pi^U_{l}$ is an
element of the $su(2)$ algebra and is purely imaginary. To understand
this, we need to notice two properties. First of all, the object
$U_l \Pi^U_l$ transforms under gauge transformations as a gauge
vector at the point $\bullet l$. The second property is that Gauss' law can be
rewritten as,
\begin{equation}
G\equiv \sum_{|l|\in L_n} \left( \kappa_{\bullet l} E_l +
\kappa_{l \bullet} E_{\overline{l}}\right)=0
\end{equation}
where $\kappa_{v}$ is a smearing function. 
This is the usual expression of Gauss' law in lattice gauge theory,
and shows that the quantity $U\Pi^U$ plays the role of an electric
field, and the Gauss law just states that the outgoing flux of
electric field lines from point $i$ vanishes.  
Conservation of Gauss' law is guaranteed, since the
action is gauge invariant, but it can be explicitly verified.

We still need to conserve in time the constraint that states that
$U\Pi^U$ is a purely imaginary element of the algebra. The resulting
equation determines the Lagrange multiplier $\alpha_{l}$.  The fact
that the Lagrange multiplier is zero implies that there will be second
class constraints. Indeed, constraints (\ref{vin3},\ref{vin4}) are
second class.

To handle the second class constraints we can introduce Dirac
brackets. The Dirac brackets become,
\begin{equation}
\left\{\left(\Pi^U_{l}\right)_A^B\,,\,
\left(U_{l'}\right)_C^D\right\} = {i \over 2 \sqrt{2}}
\delta_{ll'} \left(\delta_A^D \delta_B^C -
\left(U^\dagger_{l}\right)_{AC} \left(U_{l}\right)^{BD}\right).
\end{equation}
This relationship was derived, via an independent constructive method
by Renteln and Smolin \cite{ReSm}. Based on this Dirac bracket one can
work out the Dirac brackets of the electric field with the holonomy 
and with itself and one recovers the usual brackets that people assume
in lattice gauge theory and that approximate in the limit
$a\rightarrow 0$ the Poisson bracket in the continuum of the electric
field with a holonomy.

We now proceed to write the Dirac extended Hamiltonian, which is
achieved by imposing strongly the second class constraints and adding
the first class secondary constraints we found. Performing the sum, we
get,  
\begin{equation}
H={-1\over 2a}\sum_{|l|\in L_n} {\rm Tr}\left(
E_{l}^2+\kappa_{\bullet l} E_l + \kappa_{l \bullet}
E_{\overline{l}}\right) -{1\over 2a} \sum_{P} {\rm
Tr}\left(U_{P}\right).
\end{equation}

And we see we recover the well known Hamiltonian of Yang--Mills theory on a
lattice. This example is revealing in that if we had simply taken equation
(\ref{70}) and attempted a direct quantization ignoring the fact that there
were second class constraints it would have been very difficult to reproduce
the usual Yang--Mills Hamiltonian in the lattice. All attempts to quantize
the Hamiltonian constraint of general relativity have ignored the complex
canonical structure of the discrete Hamiltonian being quantized. To 
address this issue is the main point of our approach. 

This example also exhibits how treating time as a continuum variable while
discretizing in space yields a complicated canonical structure for the
resulting theory. In the case of general relativity it would be expected
that complications will be even greater, since the constraint algebra
intimately ties space and time. We will therefore from now on only 
discuss field theories in which both space and time are discretized.
We start by considering the example of Maxwell theory.

\subsection{Maxwell theory in a discrete space-time}

As in the case of Yang--Mills we start by writing the Wilson
action, particularized to the case of a $U(1)$ connection,
\begin{eqnarray} \label{MxwLg}
L(n,n+1) &=& -{1 \over 2} {a \over a_0} \sum_{|l|\in L_n} \left(
U^*_{l,n+1} V_{\bullet l,n}^* U_{l,n} V_{l\bullet,n} + U_{l,n+1}
V_{\bullet l,n} U^*_{l,n} V^*_{l\bullet,n} \right)
+{1 \over 2} {a_0 \over a}\sum_{P_n} U_{P_n}\nonumber\\
&&+ a_0 a^3 \sum_{|l|\in L_n} \alpha_{l,n} \left( U_{l,n}U^*_{l,n}
-1\right) +  a_0 a^3 \sum_{v\in L_n} \beta_{v,n}\left(
V_{v,n}V^*_{v,n} -1\right)
\end{eqnarray}
where as before $L_n$ es the spatial sub-lattice at a given time $n$
and we label the links with  $l$, the vertices with $v$ and the
plaquettes with $p_n$. The sum $|l|\in L_n$ goes over all links
traversed only once in a given direction. The holonomies along
spatial links are denoted by $U_{l,n}$ where $l$ is the link and $n$
the ``time level'' at which they live. Holonomies along
timelike links are denoted by $V_{v,n}$ where $v$ is the spatial
label of the vertex they start in. In this case $U$ and $V$ are
complex numbers. Given a complex number  $A$ we will denote  $A^0$ and
$A^1$ its real and imaginary parts respectively and we will write $A=A^I T^I$
where $T^0=1$ y $T^1=i$. We will take as generalized coordinates
$U^I$, $V^I$ and the Lagrange multipliers $\alpha_{l,n}$ y
$\beta_{v,n}$.

We shall now derive the canonically conjugate momenta and the equations
of motion for each variable.
For $\alpha$ and $\beta$ we get
\begin{eqnarray}
\Pi^\alpha_{l,n+1}&=& \frac{\partial L(n,n+1)}{\partial \alpha_{l,n+1}} = 0\label{MxwPiA1} \\
\Pi^\alpha_{l,n}&=& -\frac{\partial L(n,n+1)}{\partial
\alpha_{l,n}} =
-a_0 a^3 \left( U_{l,n}U^*_{l,n} -1\right) \label{MxwPiA2} \\
\Pi^\beta_{v,n+1}&=& \frac{\partial L(n,n+1)}{\partial \beta_{v,n+1}} = 0\label{MxwPiB1} \\
\Pi^\beta_{v,n}&=& -\frac{\partial L(n,n+1)}{\partial \beta_{v,n}}
= -a_0 a^3 \left( V_{v,n}V^*_{v,n} -1\right) \,.\label{MxwPiB2}
\end{eqnarray}

Equations (\ref{MxwPiA1}) y (\ref{MxwPiB1}) are primary constraints and 
their preservation leads to the secondary constraints
\begin{eqnarray}
U_{l,n}U^*_{l,n} -1 &=& 0 \label{MxwU1}\\
 V_{v,n}V^*_{v,n} -1 &=& 0\,. \label{MxwV1}
\end{eqnarray}
And these constraints are automatically preserved.

The momentum canonically conjugate to the variable
$U_{l,n+1}$ is
\begin{equation}
\Pi^U_{l,n+1} = \frac{\partial L(n,n+1)}{\partial U^I_{l,n+1}}
(T^I)^*
\end{equation}
and from the Lagrangian (\ref{MxwLg}) we get
\begin{equation}\label{MxwPiU1}
U_{l,n+1}\Pi^U_{l,n+1} = -\frac{a}{a_0} V^*_{Pl,n}
\end{equation}
where $V_{Pl,n}=U_{l,n}V_{l\bullet,n}U^*_{l,n+1}V^*_{\bullet l,n}$
is the holonomy along the plaquette that starts at the link $l$ and
continues along a timelike link. It should be noted that the
evolution equations together with the constraints that imposed
unitarity on the holonomies imply that the momenta are unitary up
to a factor,
\begin{equation}
\Pi^U_{l,n}\left(\Pi^U_{l,n}\right)^*={a^2\over a_0^2}
\end{equation}
which we will later need to impose strongly as a constraint. 

For the momentum 
$\Pi^U_{l,n}$ the Lagrangian leads to 
\begin{equation}\label{MxwPiU2}
U_{l,n}\Pi^U_{l,n} = -2a_0a^3 \alpha_{l,n} + \frac{a}{a_0}V_{Pl,n}
- \frac{a_0}{a} \sum_{P_l} U_{P_l}
\end{equation}
where  $U_{P_l}$ is the holonomy along a spatial plaquette
containing the oriented link $l$ and the sum is over all these
plaquettes. The real part of this equation determines the 
value of the variable $\alpha_{l,n}$ and the imaginary part leads
to the following evolution equation (on the constraint surface)
\begin{equation}\label{MxwPiU3}
U_{l,n+1}\Pi^U_{l,n+1} - \left(U_{l,n+1}\Pi^U_{l,n+1}\right)^* =
U_{l,n}\Pi^U_{l,n} - \left(U_{l,n}\Pi^U_{l,n}\right)^* -
\frac{a_0}{a} \sum_{P_l} \left( U^*_{P_l} - U_{P_l}\right)
\end{equation}

For the variable $V$ the canonically conjugate momenta at times $n$ 
and $n+1$ are
\begin{eqnarray}
\Pi^V_{v,n+1} &=& 0\label{MxwPiV1}
\\
V_{v,n}\Pi^V_{v,n} &=& -2a_0a^3 \beta_{v,n} + \frac{a}{a_0}
\sum_{|l|\in L_n} \left( \delta_{v,l\bullet} V_{Pl,n} +
\delta_{v,\bullet l} V^*_{Pl,n}\right) \,.\label{MxwPiV2}
\end{eqnarray}

Studying the evolution of the constraint (\ref{MxwPiV1}), the real
part of equation (\ref{MxwPiV2}) determines the value of the 
variable $\beta_{v,n}$
and the imaginary part (substituting the previous constraints)
leads to the constraint that will end up being 
Gauss' law,
\begin{equation}\label{MxwGauss}
\sum_{|l|\in L_n} \sigma_{l,v}\left[U_{l,n+1}\Pi^U_{l,n+1} -
\left(U_{l,n+1}\Pi^U_{l,n+1}\right)^* \right] =0
\end{equation}
where $\sigma_{l,v}=\delta_{v,\bullet l} -\delta_{v,l\bullet}$.
Notice also that there is no evolution equation for the $V$'s, we
will later see that they are free parameters in the theory that
correspond to gauge transformations.

Let us review the constraint structure of the theory. We have the
following constraints,
\begin{eqnarray}
U_{l,n}U^*_{l,n} &=& 1,\label{cons1}\\
V_{v,n}V^*_{v,n} &=&1,\label{cons2}\\
\Pi^U_{l,n}\left(\Pi^U_{l,n}\right)^*&=&{a^2\over a_0^2},
\label{cons3}\\
V_{v,n+1} \Pi^V_{v,n+1}+c.c.&=&0,\label{cons4a}\\
V_{v,n+1} \Pi^V_{v,n+1}-c.c.&=&0,\label{cons4b}\\
\Pi^\alpha_{l,n+1}&=&0,\label{cons5}\\
\Pi^\beta_{v,n+1}&=&0,\label{cons6}\\
\sum_{|l|\in L_n} \sigma_{l,v}\left[U_{l,n+1}\Pi^U_{l,n+1} -c.c.
\right] &=&0,\label{cons7}
\end{eqnarray}
where we have divided the constraint $\Pi^V_{v,n+1}=0$ into the
two linear combinations (\ref{cons4a},\ref{cons4b}). The set of
constraints (\ref{cons4b}-\ref{cons7}) is first class. The
remaining constraints are second class. The constraints that state
that $U$ and $\Pi$ are unitary are second class. We need to work
out the non-vanishing Dirac brackets, given by

\begin{eqnarray}
\left\{U_{l,n},\Pi^U_{l,n}\right\}_{\rm Dirac} &=& 2 -
{2\left(\Pi^U_{l,n}U_{l,n}\right)^* \over
\left(\Pi^U_{l,n} U_{l,n}\right)^* +\Pi^U_{l,n}U_{l,n}}\\[1.5ex]
\left\{U_{l,n},\left(\Pi^U_{l,n}\right)^*\right\}_{\rm Dirac} &=&
{-2\left(\Pi^U_{l,n}\right)^* U_{l,n} \over
\left(\Pi^U_{l,n}U_{l,n}\right)^* +\Pi^U_{l,n}U_{l,n}}\\[1.5ex]
\left\{V_{v,n},\Pi^V_{v,n}\right\}_{\rm Dirac} &=& 1 \\[1.5ex]
\left\{V_{v,n},\left(\Pi^V_{v,n}\right)^*\right\}_{\rm Dirac} &=&
-\left(V_{v,n}\right)^2
\end{eqnarray}
and their complex conjugate. At this point we are, in
principle, finished with the classical treatment. What we now need
to do is to perform a quantization, which entails finding the
physical space of states (states that are annihilated by the first
class constraints) and a unitary evolution operator that embodies
the finite canonical transformation that yielded the above
equations of motion.

In order to identify the evolution operator it is desirable to simplify
the structure of the theory. This can be achieved through the
introduction of real variables
\begin{eqnarray}
E_{l,n} &\equiv&  \frac{i}{2} \left(\Pi^U_{l,n} U_{l,n} -
\left(\Pi^U_{l,n}\right)^*
\left(U_{l,n}\right)^* \right),\label{96}\\
E^V_{v,n} &\equiv&   \frac{i}{2} \left(\Pi^V_{v,n} V_{v,n} -
\left(\Pi^V_{v,n}\right)^* \left(V_{v,n}\right)^* \right),
\end{eqnarray}
that have simple Dirac brackets,
\begin{eqnarray}
\left\{E_{l,n},U_{l,n}\right\}_{\rm Dirac}&=&-i\, U_{l,n} 
\label{poiseu}\\
\left\{E_{l,n},\Pi^U_{l,n}\right\}_{\rm Dirac}&=&+i\,\, \Pi^U_{l,n},\\
\left\{E^V_{v,n},V_{v,n}\right\}_{\rm Dirac}&=&-i  V_{v,n}\\
\left\{E^V_{v,n},\Pi^V_{v,n}\right\}_{\rm Dirac}&=&+i\,
\Pi^V_{v,n}.
\end{eqnarray}
In terms of these variables, the first class constraints
(\ref{cons4b}) and (\ref{cons7}) become,
\begin{eqnarray}
E^V_{v,n}&=&0,\\
\sum_{|l|\in L_n} \sigma_{l,v} E_{l,n}&=&0,
\end{eqnarray}
and we recognize in the last expression Gauss' law. The simple
canonical structure of these variables immediately suggests the
introduction of variables canonically conjugate to $E_{l,n}$ and
$E^V_{v,n}$, which we call $\varphi_{l,n}$ and $\phi_{v,n}$
respectively and are given given by,
\begin{eqnarray}
U_{l,n}&=&\exp\left(i \varphi_{l,n}\right)\\
V_{v,n}&=&\exp\left(i \phi_{v,n}\right)
\end{eqnarray}
with canonical Dirac brackets
\begin{eqnarray}
\left\{\varphi_{l,n},E_{l,n}\right\}_{\rm Dirac} &=& 1,\\
\left\{\phi_{v,n},E^V_{v,n}\right\}_{\rm Dirac} &=& 1.
\end{eqnarray}

The new variable has remarkably simple evolution equations. Starting 
from (\ref{MxwPiU3}) we get
\begin{equation}
E_{l,n+1} = E_{l,n} + i {a_0 \over 2a} \sum_{P_{l}}
\left[U(P_{l})-U^*(P_{l})\right] \,. \label{100}
\end{equation}

We need to find an evolution equation for $\varphi$. In order to
do this, we consider the definition of $E$, equation (\ref{96})
evaluated at $n+1$, and we substitute $\Pi^U_{l,n+1}$ using
equation (\ref{MxwPiU1}). The result is an alternative expression
for $E_{l,n+1}$,
\begin{equation}
E_{l,n+1}= i {a \over 2 a_0} \left(V_{Pl,n} - V_{Pl,n}^* \right).
\end{equation}
We now rewrite the
right hand side in terms of the angular variables,
\begin{equation}
E_{l,n+1}= {a \over a_0} \sin\left(\varphi_{l,n+1}-
\phi_{l\bullet,n}-\varphi_{l,n}+\phi_{\bullet l ,n}\right) \,.
\label{102}
\end{equation}
If we now equate the right hand sides of (\ref{100}) and
(\ref{102}) we get the desired evolution equation,
\begin{equation}
\varphi_{l,n+1}=\phi_{l\bullet,n}+\varphi_{l,n}-\phi_{\bullet l
,n} +\sin^{-1}\left( {a_0 \over a} E_{l,n} +i {a_0^2 \over 2a^2}
\sum_{P_{l}} \left[U(P_{l})- U^*(P_{l})\right]\right).\label{103}
\end{equation}

With the construction of the simplified variables, we are ready to
quantize the theory. We have a canonical Dirac bracket among
variables, so it is immediate to pick a representation for
operators and wavefunctions. We choose wavefunctions
$\Psi(\varphi_{l}, \phi_v,\alpha_{l},\beta_v)$.  Imposing the
first-class constraints (\ref{cons4b}-\ref{cons7}) implies that
the wavefunctions are independent of $\alpha$, $\beta$, $\phi$,
and that as functions of $\varphi$ they have to be gauge-invariant
(this is the content of the quantum Gauss law).

The quantum evolution equations are obtained by promoting all
quantities in equations (\ref{100}) and (\ref{103}) to quantum
operators. The important task is to construct a unitary evolution
operator that would embody the time evolution implied by the above
discrete quantum evolution equations. We proceed in steps. In order to
reproduce the evolution equation for $E$ we notice that,
\begin{equation}
\hat{E}_{l,n+1} = \exp\left(-i {a_0\over 2 a} \sum_{Pl}
\left[\hat{U}_{Pl}+ \hat{U}^{-1}_{Pl}\right]\right)
\hat{E}_{l,n}\, \exp\left(i {a_0\over 2 a} \sum_{Pl}
\left[\hat{U}_{Pl}+ \hat{U}^{-1}_{Pl}\right]\right) =\hat{A}
\hat{E}_{l,n} \hat{A}^{-1}
\end{equation}
where by
\begin{equation}
\hat{A} =\exp\left(-i {a_0\over 2 a} \sum_P \hat{U}_P \right)
\end{equation}
we denoted the ``partial'' evolution operator y we denote
$U_{\overline{l}} = U^{-1}_l$ . To reproduce the equation
for $\varphi$ we notice that,
\begin{equation}
\hat{\varphi}_{l,n+1} = \hat{\phi}_{l\bullet,n} +
\hat{\varphi}_{l,n} - \hat{\phi}_{\bullet l,n} + \sin^{-1}\left(
{a_0 \over a} \hat{E}_{l,n+1}\right)\label{103b}
\end{equation}
and therefore,
\begin{eqnarray}
\hat{\varphi}_{l,n+1} &=& \hat{A} \exp\left(i\sum_{|l'|\in L_n}
\left(\hat{\phi}_{\bullet l',n} - \hat{\phi}_{l'\bullet,n}\right)
\hat{E}_{l',n}\right) \left[\hat{\varphi}_{l,n} +\sin^{-1}\left(
{{a_0 \over a} \hat{E}_{l,n}} \right)\right] \nonumber
\\
&&\times\exp\left(-i\sum_{|l'|\in L_n} \left(\hat{\phi}_{\bullet
l',n} - \hat{\phi}_{l'\bullet,n}\right) \hat{E}_{l',n}\right)
\hat{A}^{-1}\nonumber
\\
&=&\hat{A} \exp\left(i\sum_{|l'|\in L_n}\left[
\left(\hat{\phi}_{\bullet l',n} - \hat{\phi}_{l'\bullet,n}\right)
\hat{E}_{l',n} +\int dx \sin^{-1}\left({a_0\over
a}x\right)\right]_{x=\hat{E}_{l',n}}\right) \hat{\varphi}_{l,n}
\nonumber
\\
&& \times\exp\left(-i\sum_{|l'|\in L_n}\left[
\left(\hat{\phi}_{\bullet l',n} - \hat{\phi}_{l'\bullet,n}\right)
\hat{E}_{l',n} +\int dx \sin^{-1}\left({a_0\over
a}x\right)\right]_{x=\hat{E}_{l',n}}\right) \hat{A}^{-1}.
\end{eqnarray}

We have therefore obtained the evolution operator,
\begin{equation}
\hat{F} = \exp\left(-i {a_0\over 2 a} \sum_P \hat{U}_P \right)
\times \exp\left(i\sum_{l'\in L_n}\left[ \hat{\phi}_{\bullet l',n}
\hat{E}_{l',n} +\frac{1}{2}\int dx \sin^{-1}\left({a_0\over
a}x\right)\right]_{x=\hat{E}_{l',n}}\right).
\end{equation}
Where the sums are over all possible plaquette and link orientations
and we are using that 
$E_{\overline{l}}=-E_l$. This
evolution operator coincides with the one obtained with the
``transfer matrix method'' in ordinary lattice Maxwell theory (the
term with the Gauss law is not recovered since the transfer matrix
method is usually applied in a gauge fixed context). See for
instance formula (5.13) of reference \cite{Creutz}. One can take
the continuum time limit of this expression $a_0\rightarrow 0$ and
one recovers the usual Hamiltonian of lattice Maxwell theory.

This concludes our discussion of ordinary lattice gauge theory. It is
evident that the proper use of discrete canonical techniques is
possible and reproduces well known results of the transfer-matrix
method. This method is geared towards theories with a definite time
variable and an associated true Hamiltonian. We will now discuss
fully constrained theories, which are less suited for treatment
with transfer-matrix methods.

\section{BF theories on the lattice}

To our knowledge, the transfer-function method has not been applied to
fully constrained field theories (like BF theory or general relativity)
without performing a gauge fixing (for gauge fixed attempts to lattice
gravity see \cite{Pelissetto}). The techniques we are introducing in
this paper, however, are not limited in their ability to handle 
fully constrained theories. This is good, since it gives hope that
they could work in the gravitational case. To begin with  we will 
consider the case of BF theory. BF theory has been quantized by many
techniques, since it has a finite number of degrees of freedom.
The interest of our approach is that it will yield a discrete theory
that has a proper canonical formulation and that contains within it
solutions that correctly approximate those of the continuum theory
at a quantum level.

The treatment we discuss applies to BF theories in any dimension. We
consider a continuum action of the form,
\begin{equation}
S = \int {\rm Tr}\left(B\wedge F\right)
\end{equation}
where $F$ is a 2-form associated with a connection (to make things
concrete, let us assume it is an $su(2)$ connection, although our
method extends easily to other semisimple groups) and $B$ is a $D-2$
$su(2)$-valued form.

In the continuum, if we perform an $N+1$ decomposition, where $N=D-1$,
one ends up with a system of first-class constraints given by,
\begin{eqnarray}
D_a E^a&=&0,\\
F_{ab}&=&0,
\end{eqnarray}
where $E^a=\epsilon^{abc\ldots} B_{bc\ldots}$ is the spatial dual of
the spatial pull-back of the form $B$ and $F_{ab}$ is the spatial
pull-back of the curvature $F$, indices $a,b,\ldots$ run from 1 to
$D-1$. So the content of the theory is that the connection is flat and
there is a Gauss law.  Quantum mechanically, the constraints are
solved by functionals of connections that have support on flat
connections only and are $SU(2)$ invariant. The theory only therefore
has a finite number of degrees of freedom and is non-trivial only in
non-trivial spatial topologies.  In $(2+1)$ dimensions the theory
(with an $SU(2)$ group) is equivalent to Euclidean Einstein gravity.

\begin{figure}[h]
\centerline{\psfig{file=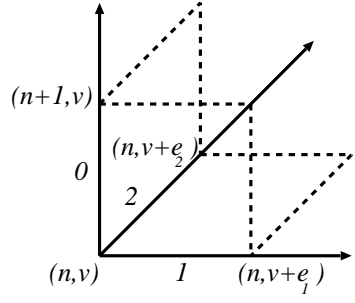,height=40mm}}
\label{fig1}
\caption{The elementary cell which we use to discretize BF theory.}
\end{figure}

We discretize the BF action in the following way,
\begin{eqnarray}
L(n,n+1)&=& \sum_{v} {\rm Tr}\left[ \rule[4ex]{0pt}{0pt} B_{n,v}^0
h^1_{n,v} h_{n,v+e_1}^2 \left(h^1_{n,v+e_2}\right)^\dag
\left(h^2_{n,v}\right)^\dag+ B^2_{n,v} V_{n,v}
h^1_{n+1,v}\left(V_{n,v+e_1}\right)^\dag
\left(h^1_{n,v}\right)^\dag\right.\\
&&-\left.B^1_{n,v} V_{n,v} h^2_{n+1,v}
\left(V_{n,v+e_2}\right)^\dag   \left(h^2_{n,v}\right)^\dag
+\mu_{n,v} \left(V_{n,v} V^\dag_{n,v}-I\right) +\sum_{k=1}^2
\lambda^k_{n,v}\left( h^k_{n,v}
\left(h^k_{n,v}\right)^\dag-I\right) \right].\nonumber
\end{eqnarray}
To explain the notation we refer to figure \ref{fig1}. From now on we
will assume we are in $2+1$ dimensions in order to simplify notation
although it will be evident that simple generalizations of the
formulae will hold in any number of dimensions. By $h^k_{n,v}$ we mean
the holonomy along the direction $k$ (in the $2+1$ dimensional case it
is either along the elementary unit vectors $e_1$ or $e_2$) starting
at the lattice point labeled by the time step $n$ and the spatial
point $v$ (in $2+1$ dimensions $v$ is labeled by a pair of
indices. By the index $v+e_k$ we mean the lattice point arrived to by
following $e_k$ starting at lattice point $v$.  The $B$ fields live
$D-2$ hypersurfaces dual to the plaquette on which we compute the
holonomy representing the $F$ and are elements of the
algebra of $SU(2)$. In $2+1$ dimensions $B$ is a one-form
we therefore write it with one index.  The vertical holonomies
$V_{n,v}$ are only labeled by the lattice point they start
at. We will assume that the holonomies are matrices of the form
$h=h^I T^I$, $V=V^I T^I$ 
where $T^0=I/\sqrt{2}$ and 
$T^a=-i\sigma^a/\sqrt{2}$ where $\sigma_a$ are the Pauli matrices.
From now on when we take variations we will assume that one has
performed the variations with respect to the components $h^I$
and reconstituted the remaining equations as matrix equations.
$\lambda$ and $\mu$ are Lagrange multipliers that enforce, given
the above condition of the holonomies,  that they
are elements of $SU(2)$.

The fundamental canonical variables of the theory are therefore 
$h^{k,I}_{n,v}$, $V^I_{n,v}$, $B^{k,a}_{n,v}$,
$\lambda^k_{n,v}$ and $\mu_{n,v}$. The equations of motion for the variables 
$\lambda^k_{n,v}$ are,
\begin{eqnarray}
\Pi^{\lambda^k}_{n+1,v}&=& \frac{\partial L(n,n+1)}{\partial \lambda^k_{n+1,v}} = 0 \label{pil}\\
\Pi^{\lambda^k}_{n,v}&=& -\frac{\partial L(n,n+1)}{\partial
\lambda^k_{n,v}} = -{\rm Tr}\left[h^k_{n,v}
\left(h^k_{n,v}\right)^\dag-I \right]
\end{eqnarray}

Preservation in time of the primary constraint (\ref{pil}) leads to the
secondary constraint
\begin{equation} \label{hsu2}
{\rm Tr}\left[h^k_{n,v} \left(h^k_{n,v}\right)^\dag-I \right]=0
\end{equation}
and preserving in time this constraint does not yield any new
constraints.
The equations of motion for the variables 
 $\mu_{n,v}$ are
\begin{eqnarray}
\Pi^\mu_{n+1,v}&=& 0 \label{pim}\\
\Pi^\mu_{n,v}&=&  -{\rm Tr}\left[V_{n,v}
\left(V_{n,v}\right)^\dag-I \right].
\end{eqnarray}

Preserving in time constraints(\ref{pim}) leads to the constraints
\begin{equation}\label{vsu2}
{\rm Tr}\left[V_{n,v} \left(V_{n,v}\right)^\dag-I \right]=0
\end{equation}
which is preserved in time without introducing any new constraints. 

The equations of motion for the variables
$B^{k,a}_{n,v}$ are
\begin{eqnarray}
\Pi^{B^k}_{n+1,v}&=& \frac{\partial L(n,n+1)}{\partial
B^{k,a}_{n+1,v}} \,\left(T^a\right)^\dag = 0 \label{piB}\\
\Pi^{B^k}_{n,v}&=& -\frac{\partial L(n,n+1)}{\partial
B^{k,a}_{n,v}} \,\left(T^a\right)^\dag = -{\rm
Tr}\left[h^{i\,j}_{n,v} \,T^a \right] \,\left(T^a\right)^\dag
\end{eqnarray}
where $\{i,j,k\}$ is a cyclic permutation of $\{0,1,2\}$ and
$h^{i\,j}_{n,v}$ is the holonomy starting from $n,v$ around the
elementary plaquette in the plane $i-j$. Preserving in time the 
constraint (\ref{piB}) together with constraints (\ref{hsu2},\ref{vsu2})
leads to
\begin{equation}\label{hij}
h^{i\,j}_{n,v} = \sigma^{i\,j}_{n,v} \,I
\end{equation}
with $\sigma^{i\,j}_{n,v}=\pm 1$. This sign is arbitrary and can change
from plaquette to plaquette.

The equations of motion for the variables
$V^I_{n,v}$ are
\begin{eqnarray}
\Pi^V_{n+1,v}&=& \frac{\partial L(n,n+1)}{\partial
V^I_{n+1,v}} \,\left(T^I\right)^\dag = 0 \label{piV1}\\
V_{n,v}\Pi^V_{n,v}&=&  h^{0\,1}_{n,v} B^2_{n,v} - h^{0\,2}_{n,v}
B^1_{n,v} - \left(h^1_{n,v-e_1}\right)^\dag h^{1\,0}_{n,v-e_1}
B^2_{n,v-e_1} h^1_{n,v-e_1}  \nonumber\\
&& +\left(h^2_{n,v-e_2}\right)^\dag h^{2\,0}_{n,v-e_2}
B^1_{n,v-e_2} h^2_{n,v-e_2} + 2\mu_{n,v}\,I \,. \label{piV2}
\end{eqnarray}

Preserving in time the constraints (\ref{piV1}), together with (\ref{hij})
leads to determining the multipliers
\begin{equation}
\mu_{n,v} = 0
\end{equation}
and the equation of motion,
\begin{equation}
h^{0\,1}_{n,v} B^2_{n,v} - h^{0\,2}_{n,v} B^1_{n,v} -
\left(h^1_{n,v-e_1}\right)^\dag h^{1\,0}_{n,v-e_1} B^2_{n,v-e_1}
h^1_{n,v-e_1} +\left(h^2_{n,v-e_2}\right)^\dag h^{2\,0}_{n,v-e_2}
B^1_{n,v-e_2} h^2_{n,v-e_2} = 0 \,.\label{gauss1}
\end{equation}

The equation of motion for 
$\Pi^{h^1}_{n+1,v}$ is
\begin{equation}\label{pih1}
h^1_{n+1,v}\Pi^{h^1}_{n+1,v} = V^\dag_{n,v}h^{0\,1}_{n,v}
B^2_{n,v}V_{n,v}
\end{equation}
and using (\ref{hij}) we obtain the primary constraint
\begin{equation}\label{ligh1}
{\rm Tr}\left[ h^1_{n+1,v}\Pi^{h^1}_{n+1,v} \right]=0 \,.
\end{equation}
Preserving in time this constraint, together with the equation 
of motion for 
$\Pi^{h^1}_{n,v}$ leads to the fixing the multiplier
\begin{equation}
\lambda^1_{n,v} = 0
\end{equation}
and the equation of motion
\begin{equation}\label{pih2}
h^1_{n,v}\Pi^{h^1}_{n,v} = h^{1\,0}_{n,v} B^2_{n,v}
-h^{1\,2}_{n,v} B^0_{n,v} +\left(h^2_{n,v-e_2}\right)^\dag
h^{2\,1}_{n,v-e_2}B^0_{n,v-e_2}h^2_{n,v-e_2} \,.
\end{equation}

We proceed in a similar fashion for the variable $h^{2,I}_{n,v}$. The 
equations of motion lead to 
\begin{eqnarray}
h^2_{n+1,v}\Pi^{h^2}_{n+1,v} &=& -V^\dag_{n,v}h^{0\,2}_{n,v}
B^1_{n,v}V_{n,v} \\
{\rm Tr}\left[ h^2_{n+1,v}\Pi^{h^2}_{n+1,v} \right]&=&0 
\label{ligh2}\\
\lambda^2_{n,v} &=& 0 \\
h^2_{n,v}\Pi^{h^2}_{n,v} &=& h^{2\,1}_{n,v} B^0_{n,v}
-h^{2\,0}_{n,v} B^1_{n,v} -\left(h^1_{n,v-e_1}\right)^\dag
h^{1\,2}_{n,v-e_1}B^0_{n,v-e_1}h^1_{n,v-e_1} \,.
\end{eqnarray}

We now consider the variable,
\begin{equation}
E^k_{n,v} \equiv h^k_{n,v} \Pi^k_{n,v}  \quad k\in\{1,2\} \,,
\end{equation}
that our experience with Yang--Mills shows should play the role of
an electric field along the direction $e_k$ at the point
$\{n,v\}$. Due to the constraints (\ref{ligh1}) y (\ref{ligh2}),
the variable in question is an element of the 
algebra, i.e., is  a traceless matrix,
\begin{equation}
{\rm Tr}\left( E^k_{n,v}\right)=0.\label{traceE}
\end{equation}

We also define the electric field in the direction 
$-e_k$ at the point  $\{n,v\}$ by
\begin{equation}
E^{\overline{k}}_{n,v} \equiv h^{\overline{k}}_{n,v}
\Pi^{\overline{k}}_{n,v} = (h^k_{n,v-e_k})^\dag
(\Pi^k_{n,v-e_k})^\dag = -(h^k_{n,v-e_k})^\dag E^k_{n,v-e_k}
h^k_{n,v-e_k}
\end{equation}

Using the equations of motion and constraint it is straightforward
to get 
\begin{eqnarray}
E^1_{n+1,v} &=& +V^\dag_{n,v}h^{0\,1}_{n,v} B^2_{n,v}V_{n,v} \\
E^2_{n+1,v} &=& -V^\dag_{n,v}h^{0\,2}_{n,v} B^1_{n,v}V_{n,v} \\
E^{\overline{1}}_{n+1,v} &=& -V^\dag_{n,v} (h^1_{n,v-e_1})^\dag
h^{0\,1}_{n,v-e_1} B^2_{n,v-e_1} h^1_{n,v-e_1} V_{n,v} \\
E^{\overline{2}}_{n+1,v} &=& +V^\dag_{n,v} (h^2_{n,v-e_2})^\dag
h^{0\,2}_{n,v-e_2} B^1_{n,v-e_2} h^2_{n,v-e_2} V_{n,v}
\end{eqnarray}
and from here one can show that  (\ref{gauss1}) is
equivalent to
\begin{equation}\label{gauss2}
E^1_{n+1,v} + E^2_{n+1,v} + E^{\overline{1}}_{n+1,v} +
E^{\overline{2}}_{n+1,v} =0 \,.
\end{equation}
This is the familiar expression for Gauss' law on the lattice.

We have noted that the holonomy along the plaquette $1-2$ is
proportional to the identity. Therefore one is dealing with a flat
connection on the plaquette. As expected, the discrete theory admits
more solutions than the continuum one, in the sense that a possible
solution would be a connection that makes the holonomy $+1$ on certain
plaquettes and $-1$ on others. Such solutions will not admit a
continuum counterpart.

We need to check that the Dirac brackets of the elementary variables
lead to the expected results. Here we can just refer to the discussion
we did in the Yang--Mills case, since the structure of the second
class constraints (\ref{hsu2},\ref{traceE}) is similar to the one
we encountered in that case.

The quantization of the theory is now immediate. The Dirac bracket
between the electric field and the holonomy is similar to the one in
the Maxwell case (\ref{poiseu}),
\begin{equation}
\left\{\left(E^k_{n,v}\right)^A,h^k_{n,v}\right\} = iT^A h^k_{n,v}
\end{equation}
where we have introduced the component notation in the algebra and
the $T^A$'s are the generators of $su(2)$ we introduced before.

One can now consider wavefunctions that are functions of the
holonomies on the links of the lattice, $\Psi[h^k_{n,v}]$. Following
the Dirac method we need to impose the constraints as operator
equations.  Gauss' law will just state that the wavefunctions are
gauge invariant. In this case it will imply that the wavefunctions are
linear combinations of traces of holonomies along closed loops.

We now need to impose the spatial projection of the constraint 
(\ref{hij}) as an operatorial
equation, 
\begin{equation}
h^{1\,2}_{n,v} = \sigma^{1\,2}_{n,v} \,I.
\end{equation}

In the space of functions of holonomies the equation becomes
multiplicative and the solution is that the holonomies along all
elementary plaquettes should be $\pm 1$ times the identity. We
therefore see that we recover as solutions of the theory the space of
gauge invariant functions of a flat connection. Except that in the
discrete theory it could happen that in some plaquettes the holonomies
are $+1$ times the identity whereas in others the holonomies have a
negative sign. Such solutions are possible in the discrete theory and
do not have a counterpart in the continuum. We see that the discrete
theory therefore has considerably richer solution structure than the
continuum theory. If one however imposes that the solutions be
continuous from one plaquette to another, one is only left with the
traditional solution.

The unitary operator that implements the finite canonical
transformation is, for this theory, given by an exponential of a
combination of the first class constraints with arbitrary
coefficients. 

\section{Conclusions}

We have generalized classical mechanics to systems in which space and
time are discrete. Time evolution is implemented through a finite
canonical transformation. Constrained systems can be incorporated in
the scheme. We discussed in detail how to handle Yang--Mills theory, 
Maxwell theory and BF theory and encountered no problem setting up the
classical and quantum theory. In general, the discrete theories 
contain many solutions that do not have a good continuum limit,
as we have seen in the BF example, and the correct solution has
to be chosen by requesting that it have the appropriate semiclassical
limit.

The framework we have set up is completely ready to be applied to the
case of general relativity, which will have many similarities with the
BF case. An attractive advantage is that the formalism does not
require an Euclidean rotation and operates directly in terms of the
Lorentzian signature theory. We will discuss the case of general
relativity in a forthcoming publication. Having a sound theoretical
foundation for dealing with discretized gravity will allow quantize
the theory in a systematic way and will enable us to make connections
between the canonical and path integral approaches.

\acknowledgements 

We wish to thank Abhay Ashtekar, Martin Bojowald and  Karel Kucha\v{r} 
for comments. This work was supported in part by grants
NSF-PHY0090091, funds from the Fulbright commission in Montevideo and
by funds of the Horace C. Hearne Jr. Institute for Theoretical
Physics.

\end{document}